\documentclass[10pt,a4paper,final,journal]{IEEEtran}

\usepackage[T1]{fontenc}
\usepackage[latin9]{inputenc}
\usepackage[active]{srcltx}
\usepackage{array}
\usepackage{multirow}
\usepackage{blindtext}
\usepackage{multicol}
\usepackage{amsmath}
\usepackage{amssymb}
\providecommand{\keywords}[1]{\textbf{\textit{Index terms}} }
\usepackage{graphicx}
\usepackage{setspace}
\usepackage{lipsum}
\usepackage[T1]{fontenc}
\usepackage{authblk}
\usepackage{url}
\usepackage{algorithm}
\usepackage{float}
\restylefloat{table}
\usepackage{xcolor}
\usepackage{multirow}
\usepackage[noend]{algpseudocode}
\makeatletter

\usepackage{cite}\usepackage{threeparttable}

\usepackage{picinpar}

\usepackage{psfrag}\usepackage{placeins}\usepackage{times}
\usepackage{amsfonts}\usepackage{amsfonts}
\usepackage{epstopdf}

\newcommand*{\rom}[1]{\expandafter\@slowromancap\romannumeral #1@}

\input epsf
\usepackage{babel}\addto\captionsbreton{}
\addto\captionsbulgarian{}
\addto\captionsenglish{}

\usepackage{babel}\addto\captionsbreton{}
\addto\captionsbulgarian{}
\addto\captionsenglish{}

\usepackage{babel}\addto\captionsbreton{}
\addto\captionsbulgarian{}
\addto\captionsenglish{}

\@ifundefined{showcaptionsetup}{}{%
 \PassOptionsToPackage{caption=false}{subfig}}
\usepackage{subfig}
\makeatother

\usepackage{babel}
\begin{document}

\title{Analogue Radio over Fiber aided Multi-service Communications for High Speed Trains}

\author{
\IEEEauthorblockN{Yichuan Li, Salman Ghafoor, Mohammed El-Hajjar, \textit{Senior Member, IEEE}}
\\

\vspace{-1cm}

\thanks{Y.  Li is with Harbin Institute of Technology (Shenzhen), Shenzhen, China (e-mail: liyichuan@hit.edu.cn).}
\thanks{S. Ghafoor is with the National University of Sciences and Technology,  Pakistan (e-mail: salman.ghafoor@seecs.edu.pk).}
\thanks{M. El-Hajjar is with the School of Electronics and Computer Science, University of Southampton, SO17 1BJ, United Kingdom (meh@ecs.soton.ac.uk).}}
\maketitle
\begin{abstract}High speed trains (HST) have gradually become an essential means of transportation, where given our digital world, it is expected that passengers will be connected all the time. More specifically, the on-board passengers require fast mobile connections, which cannot be provided by the currently implemented cellular networks. Hence, in this article, we propose an analogue radio over fiber (A-RoF) aided multi-service network architecture for high-speed trains, in order to enhance the quality of service as well as reduce the cost of the radio access network (RAN). The proposed design can simultaneously support sub-6GHz as well as milimeter wave (mmWave) communications using the same architecture. Explicitly, we design a photonics aided beamforming technique in order to eliminate the bulky high-speed electronic phase-shifters and the hostile broadband mmWave mixers while providing a low-cost RAN solution. Finally, a beamforming range of 180$^\circ$ is demonstrated with a high resolution using our proposed system.\end{abstract}


\section{Introduction}
\label{Section:Introduction}
\IEEEPARstart{T}{he} high speed train (HST) operating at a speed above 300 km/hour has fundamentally changed the individuals' life-style. It is reported that HST has the second-highest internet streaming after household environments, such as homes and offices \cite{8438925,dang2007radio}, where the majority of this streaming is used for supporting the on-board activities, such as on-demand video, online-gaming, and voice or video calling. Furthermore,  since the 4G and 5G use the sub-6GHz band and then the millimeter wave (mmWave) communications will coexist with these frequency bands \cite{osseiran20165g},  a railway network supporting multi-service is required,  where the passengers are capable of being connected to 4G,  5G, Beyond 5G or even Wi-Fi networks all the time \cite{8864000,osseiran20165g,8481549}.  On the other hand, HST requires high capacity due to the increasingly high number of passengers and the severe Doppler effects caused by the high speed \cite{7968457,goldsmith2005wireless}. As a result, this requires more base-stations and an efficient cellular handover technique, which has several challenges \cite{7968457,goldsmith2005wireless,6191306,7295468}. First, the denser deployment of base stations (BS) can provide seamless connection for the passengers by increasing the capacity per cell, which by contrary would impose more inter-cell interference (ICI) \cite{7968457,goldsmith2005wireless}. Then, the severe Doppler effects seriously undermine the system as the train is moving extremely fast, impacting the handover process of the cellular networks \cite{8864000,8481549}. Besides, the channel estimation might not be accurate due to the rapid channel variations, influencing the quality of service (QoS) and hence, the user experience \cite{6191306,7295468}.
\begin{figure*}[h]
\centering
	\includegraphics[width=\textwidth]{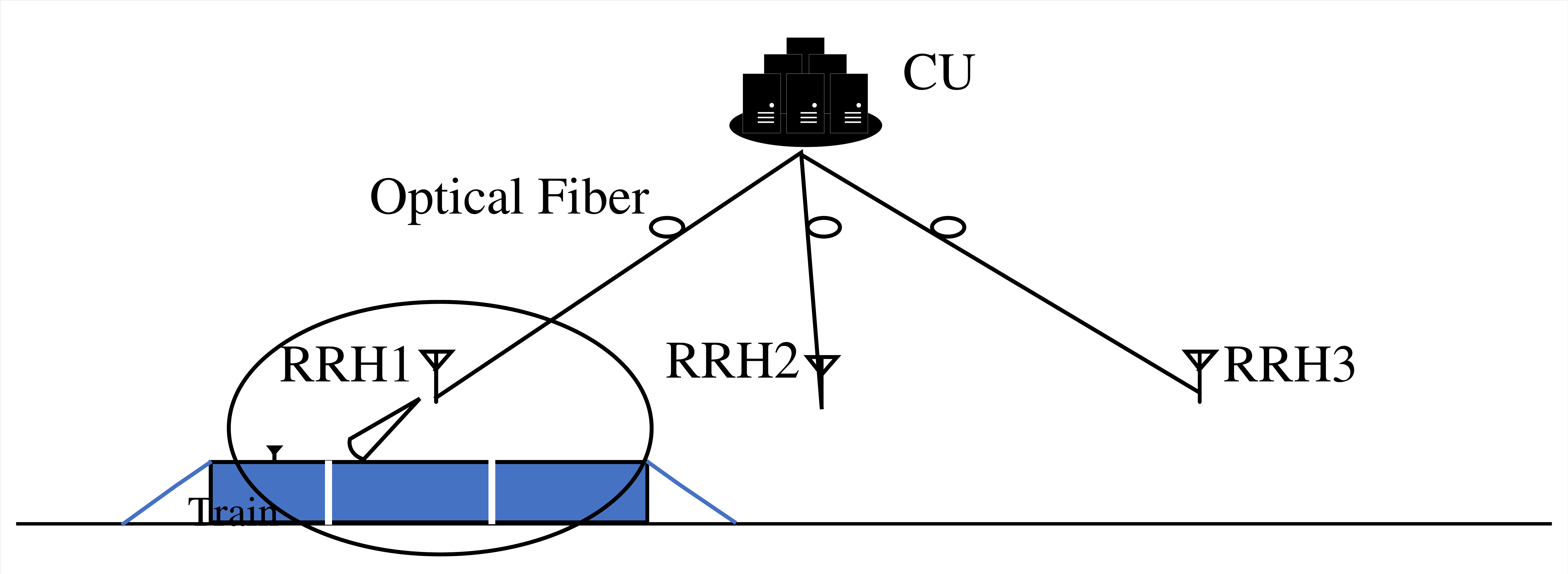}
		\caption{Multi-service HST System Architecture}
		\label{Ralay Figure}
\end{figure*}

As a solution, mmWave beamforming has been proposed for mitigating the ICI,  while enhancing the capacity,  especially in ultra-dense cellular networks \cite{8207426}. On the other hand, two-hop relay communications are advocated by \cite{8438925, 8864000} to address the handover failure and the rapid channel variation in HSTs, where several relay antennas can be installed on the rooftop of the train cars for a stable handover. However, in order to have a decent quality of service (QoS), a large number of base-stations must be deployed, increasing the total cost of the radio access networks (RANs) including both the Capital Expenditure (CAPEX) and the Operational Expenditure (OPEX). Hence, ultra-light RANs are required \cite{6897914}. In addition, mmWave communications beamforming requires high-speed mmWave phase-shifters \cite{6979963}, which increases the overall RAN cost and limits the tuning carrier frequency range \cite{7055229,9356530,cao2016advanced}, and the broadband mixers, which affects achievable the bit error rate (BER) performance \cite{7055229}. Furthermore, the phase-shifter based beamforming would impose beam-squinting on the wideband signals\footnote{Beamsquinting is the beam-shifts caused by the frequency shifts when applied with the constant phase-shift among neighbouring antenna element (AE). This would be even severe in the context of wide-band signal beamsteering \cite{cao2016advanced}.}. 

The analogue radio over (A-RoF) based true-time delay is a low-cost yet high-performance RAN solution with ultra-light remote radio head (RRH) \cite{Li:18,7968457,8328821,7017473,8616905,9356530,vasileiou2016effective}, where the RAN is separated between the central unit (CU) and several RRHs and is capable of reducing both its power-consumption and complexity. Explicitly, the A-RoF aided beamformer was proposed for providing beam-squinting free beamforming using the uniform fiber Bragg grating (FBG) \cite{cao2016advanced,molony1996fiber,ball1994programmable} or a single chirped FBG (CFBG) \cite{hunter2006demonstration,yao2002continuous}.
\begin{figure*}[h]
\centering
\includegraphics[width=.9\textwidth]{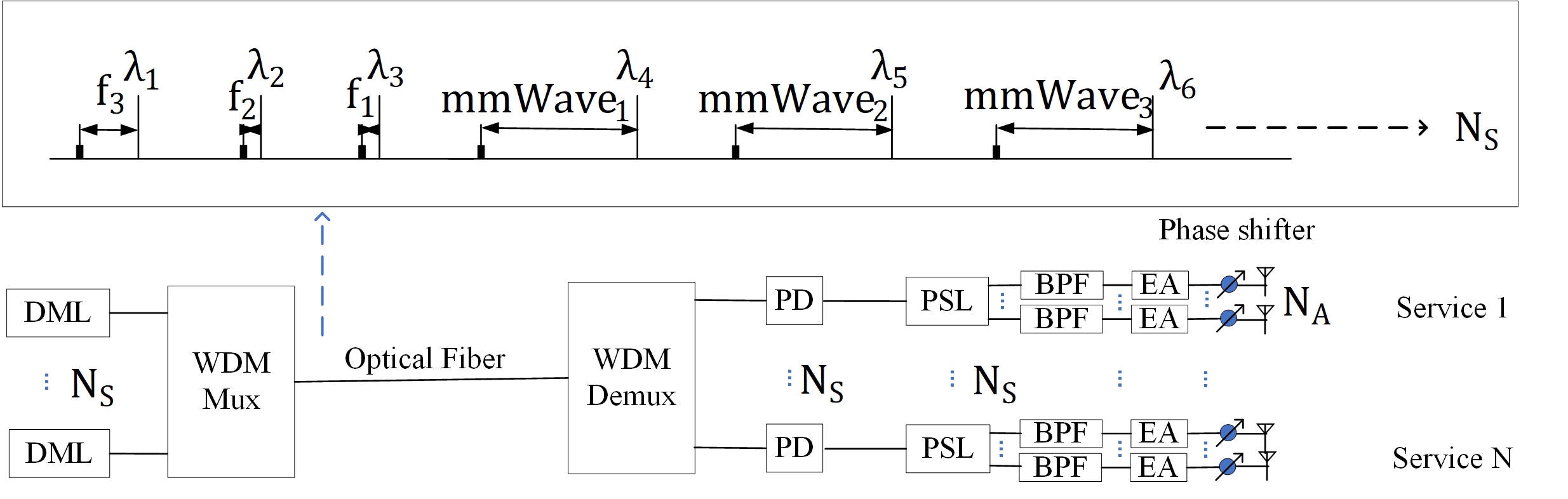}
\caption{Conventional System Model (DML: Directly Modulated Laser; WDM Mux: Wavelength Division Multiplexer; PD, Photo-detector; PSL: Power Splitter; BPF: Bandpass Filter; EA: Electronic Amplifier; NS: Number of Services; NA: Number of Antenna Element.)}
\label{Figure: Conventional System}
\end{figure*}
			
In this article, we design a two-hop HST communications network combined with the A-RoF aided beamforming techniques, where multi services, such as on-demand video, online-gaming, and voice or video calling, which are carried by mmWave and sub-6GHz signals using 4G, 5G or Wi-Fi networks are delivered to the train. Explicitly, the multi service signals generated in the CU of Fig. \ref{Ralay Figure} are transmitted through a short-length fiber of upto 20km to the RRHs, which communicate with the antennas on the top of the HST to provide the required service to the on-board passengers. The beam pattern is controlled in the CU using the CFBG, instead of the costly and bulky phase-shifters in the RRHs, alleviating the size and cost of the RRH and facilitating the densely deployed RRHs along the railways. The proposed system enables both cost-reduction by simplifying the RAN design and performance-improvement by exploiting centralised optical processing and multi-service signal generation as a benefit of the A-RoF architecture \cite{9356530}. The novel contributions of our system are as follow: 
\begin{enumerate}
\item Photonic aided true-time delay beamforming: We conceive a multi-service true-time-delay transmit beamforming for HST two-hop network, while also eliminating the beam-squinting phenomenon.
\item Energy-efficient RRHs: The large-scale power-thirsty phase shifters and costly mmWave phase shifters are removed, while the beam is optically controlled using the passive FBGs in the proposed design, hence improving the OPEX-savings and enabling ultra-light RRHs deployment.
\item Multi-service signal generation: multi-service signals carried by mmWave and sub-6GHz frequency band can be optically generated and beam-steered to the HST in the same direction.
\item Centralised optical processing: unlike the phase-shifter used in conventional wireless transmit beamforming, the beam control module is located in the central unit, potentially facilitating the coordinated multipoint (CoMP) transmission.
\end{enumerate}
The rest of the paper is organised as follows. We introduce the multi-service two-hop train system in Section \ref{Section:Multi-service Train System Architecture}, while the A-RoF aided beamforming technique and its time delay principle are detailed in Section \ref{Section:A-RoF aided Beamforming for High-speed Train Multi-service Communications}. Then, we present the cost-benefits and the beamforming performance of our proposed system in Section \ref{Section:System Evaluation}, followed by the conclusion in Section \ref{subsection: Conclusion}.

\begin{figure*}[h]
\centering
	\includegraphics[width=\textwidth]{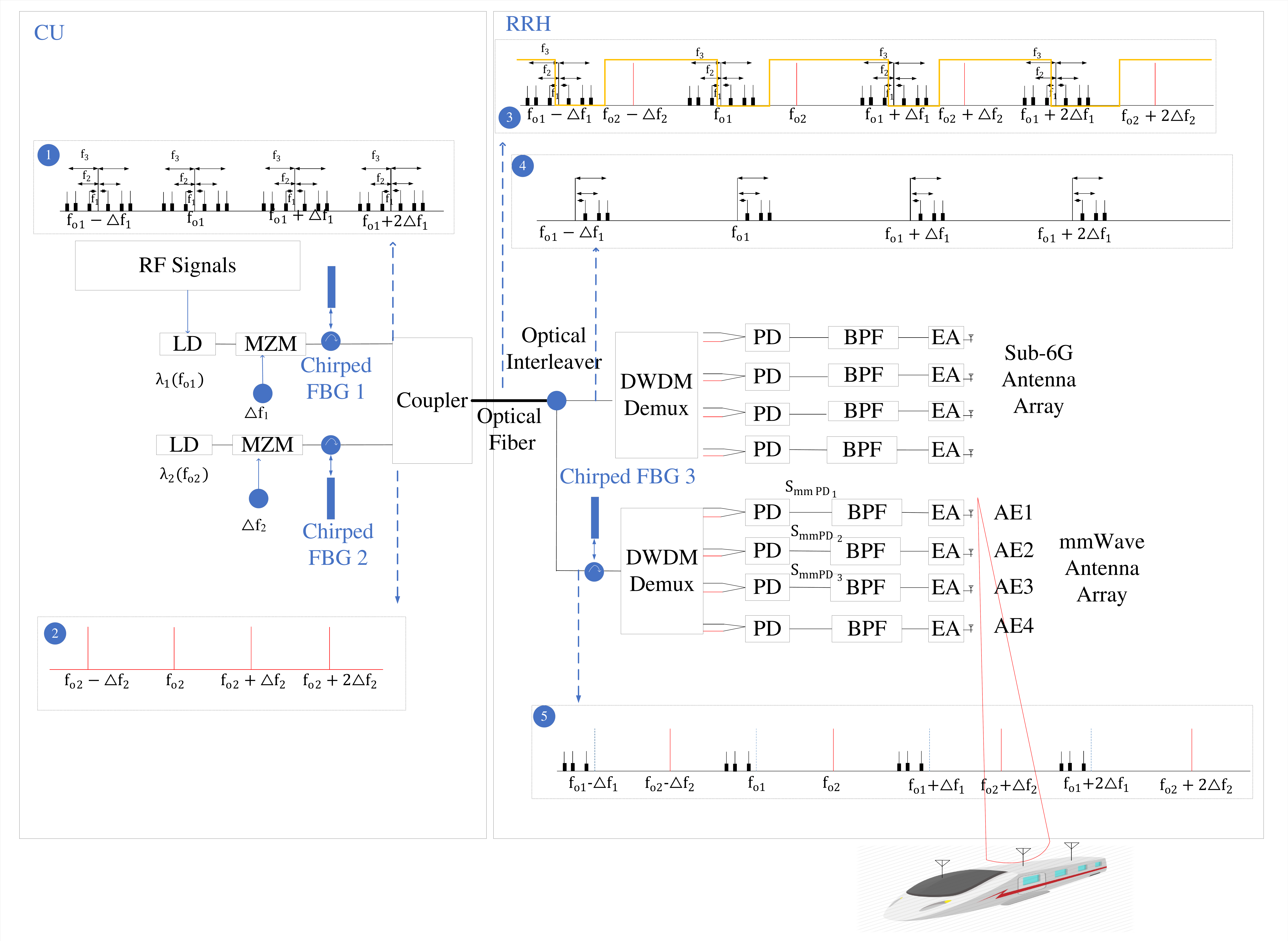}
		\caption{Proposed A-RoF aided Multi-service Architecture.}
		\label{Fig. A-RoF aided Multi-service Architecture}
\end{figure*}
		
\section{Multi-service HST System Architecture}
\label{Section:Multi-service Train System Architecture}

In this section, we present a general architecture for the two-hop HST system shown in Fig. \ref{Ralay Figure},  which can be exploited in our design.  As shown in Fig. \ref{Ralay Figure}, the signal is generated in the CU and transmitted via fiber to several RRHs, where only optical-to-electronic conversion, amplification and filtering are performed, thus substantially reducing the RRH size. Explicitly, the RRH receives the signal from the CU using fiber and then transmits this signal to the HST using a set of antenna arrays, as shown in Fig.  \ref{Ralay Figure}, where the RRHs are placed along the train line and will communicate with the antenna arrays on the rooftop of the HST.
	
The signals transmitted from the RRHs to the HST are beamformed and transmitted to the antenna arrays fixed on the rooftop of the HST, as shown in Fig. \ref{Ralay Figure}. Then, the signals are relayed to the on-board passengers to fulfill the users demands through the indoor communication system \cite{6180090}. It has been verified using field tests that the two-hop relay communication in HST setup outperforms the conventional 4G cellular system in terms of signal quality \cite{6180090}.

Since the 4G and 5G signals would co-exist in the non-standalone 5G deployment proposed by the 3GPP in the recent Releases \cite{9269939}, multi-service signal transmission using different frequency bands and wireless standards are of importance to support the users' daily streaming, on-line gaming, voice or video call and on-demand video \cite{osseiran20165g}. 

Conventionally, the A-RoF aided multi-service system can be supported using the architecture of Fig. \ref{Figure: Conventional System} \cite{8438925, 8864000, 7295468, 7054728, 6180090}, where different services carried by different frequencies are directly modulated by individual lasers before being coupled into an optical fiber. At the end of the fiber, each optical and radio frequency (RF) chain are used for optical-to-electronic conversion, amplification, filterings. Furthermore, when analogue beamforming is employed for improving the signal-to-noise ratio performance, a large number of phase-shifters, band-pass filter (BPFs), electronic amplifiers (EAs) are required, which increases the cost and power-consumption of the whole system. Explicitly, let us take service 1 as an example, as shown in Fig. \ref{Figure: Conventional System}, assuming the RF of $f_3$ carries service 1, $f_3$ is used to directly modulate a laser operating at the center wavelength of $\lambda_1$. Then, coupled with other services carried by $\lambda_2$ to $\lambda_6$ into the fiber, the wavelength division multiplexing (WDM) signal of the spectrum shown in Fig. \ref{Figure: Conventional System} feeds the optical fiber for transmission, after which the WDM Demultiplexer (WDM Demux) separates each wavelength. Then, the modulated optical signal of $\lambda_1$ is photo-detected and the $f_3$ carrying service 1 is recovered, followed by the power splitter (PSL) which splits the power of the RF frequency, where, each BPF and EA is used for filtering out the wanted signal and for the amplification of each port of the PSL, respectively. Finally, several phase-shifters are used to phase-shift each output to form a desired beam pattern, leading to the analogue beamforming with large amount of power-thirsty and bulky phase-shifter deployment. In the next section, we will introduce our A-RoF system based on the two-hop transmission, while invoking a lower-cost photonic aided multi-service solution.
		\begin{figure*}[h]
\centering
	\includegraphics[width=\textwidth]{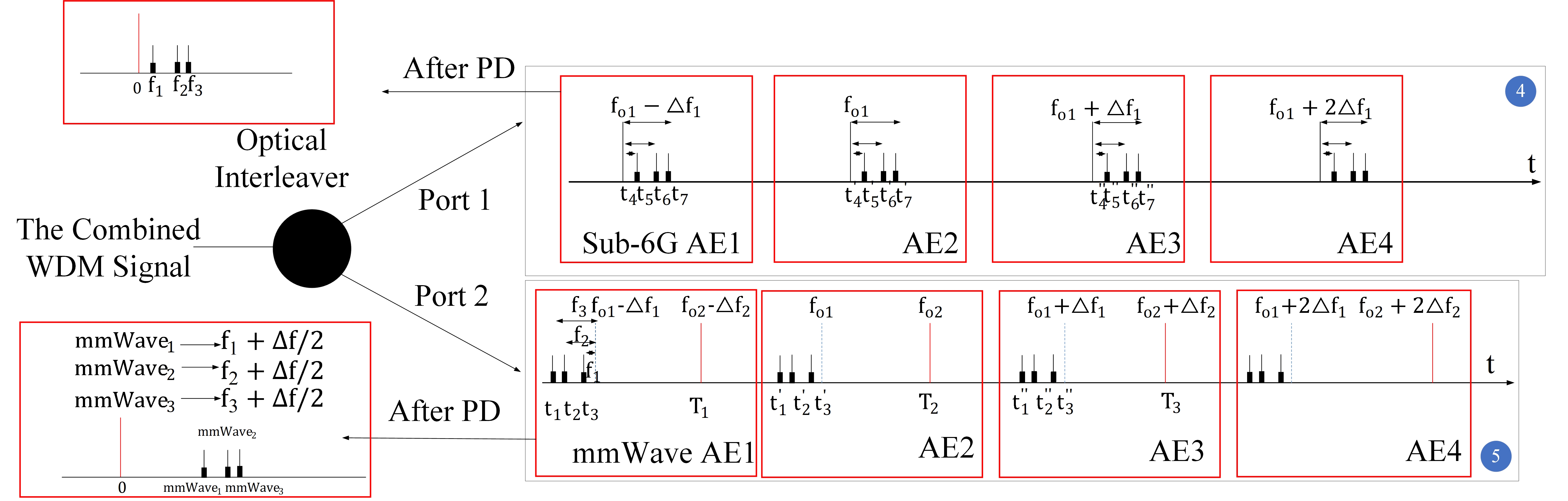}
		\caption{The proposed network's multi-service signal generation and their time-delay imposed.}
		\label{Figure: The sub-6G and mmWave generation and their time-delay imposed}
		\end{figure*}
\section{A-RoF aided Beamforming for High-speed Train Multi-service Communications}
\label{Section:A-RoF aided Beamforming for High-speed Train Multi-service Communications} 
\subsection{A-RoF aided HST System Model}
As mentioned in section \ref{Section:Multi-service Train System Architecture}, the RRH beamforms the multi-service signals, composed of sub-6GHz and mmW signals, to the HST rooftop antenna arrays. Fig. \ref{Fig. A-RoF aided Multi-service Architecture} shows the proposed A-RoF design, where the sub-6GHz and mmW signals are beamformed in the same direction to the antenna array placed on the rooftop of the train. The proposed design avoids the beam-squinting as well as considers low-complexity RRH. 

In Fig. \ref{Fig. A-RoF aided Multi-service Architecture}, the access network is divided into CU and RRH, where the CU performs the baseband signal processing as well as the radio modulations \cite{9356530}, while the RRH is simplified to only radio functions such as amplifications and filtering. Explicitly, in the CU of Fig. \ref{Fig. A-RoF aided Multi-service Architecture}, the RF signals of $f_1$, $f_2$ and $f_3$\footnote{The number of frequencies can be readily extended, which depends on the bandwidth of the LD.} directly modulate a laser diode (LD) operating at $\lambda_{1}$. The signal at the output of the directly modulated laser is given at the optical input of a Mach-Zehnder modulator (MZM) that is being driven by a sinusoidal signal of frequency $\Delta f1$. The nonlinear transfer function of the MZM results in the generation of multiple spectral copies of the directly modulated optical signal that have a frequency spacing of $\Delta f_{1}$ as shown in Fig. \ref{Fig. A-RoF aided Multi-service Architecture} \cite{7055229}. The resulting WDM signal is time-delayed using a CFBG, which is capable of reflecting different wavelengths with a linear time-delay. Then, as shown in Fig. \ref{Fig. A-RoF aided Multi-service Architecture}, another laser diode operating at $\lambda_2$ is intensity modulated by a sinusoidal signal having frequency of $\Delta f_{2}$ through a MZM to generate multiple sidebands by using the nonlinear function of the MZM \cite{7055229}. The two WDM signals are coupled into a single mode fiber using a fiber coupler. 

As shown in the RRH of Fig. \ref{Fig. A-RoF aided Multi-service Architecture}, the combined WDM signal is separated by a two-port optical interleaver, which filters the desired signals to two individual dense wavelength division demultiplexers (DWDM Demux), each of which would be used for either sub-6GHz signal generation or mmWave signal generation. The upper DWDM Demux of Fig. \ref{Fig. A-RoF aided Multi-service Architecture} filters the spectrum which can generate a beat frequency of sub-6GHz bands. Then, the photodetected sub-6GHz frequency after each PD is band-pass filtered and amplified, while the inter-element time delay is dependent on the chirped FBG 1 and FBG 2 strains as tested by \cite{1022008}. Similarly, the bottom DWDM Demux in the RRH of Fig. \ref{Fig. A-RoF aided Multi-service Architecture} is used for mmWave signal generation, where an extra chirped FBG, namely chirped FBG 3 of Fig. \ref{Fig. A-RoF aided Multi-service Architecture}, is implemented for adjusting the beam direction to be the same as the sub-6GHz bands, which will be elaborated in the next section. 

To elaborate further on the RF and mmWave signals generation, as shown in Fig. \ref{Figure: The sub-6G and mmWave generation and their time-delay imposed}, the optical interleaver, which is a periodic optical filter, separates the combined WDM signals of spectrum \textcircled{3} of Fig. \ref{Fig. A-RoF aided Multi-service Architecture}. The spectrum of the two outputs of the optical interleaver is depicted in Fig. \ref{Figure: The sub-6G and mmWave generation and their time-delay imposed}, where port 1 maps the spectrum \textcircled{4} of Fig. \ref{Figure: The sub-6G and mmWave generation and their time-delay imposed} to the sub-6GHz spectrum of $f_1$, $f_2$ and $f_3$, with port 2 mapping the spectrum \textcircled{5} to the mmWave frequencies of $f_1+\frac{\Delta f}{2}$, $f_2+\frac{\Delta f}{2}$ and $f_3+\frac{\Delta f}{2}$\footnote{Note that if $\frac{\Delta f}{2}=f_{o1}-f_{o2}$ is upto 25 GHz, $f_1+\frac{\Delta f}{2}$ is the mmWave signal of at least 25 GHz. The spectrum \textcircled{4} and \textcircled{5} of Fig. \ref{Figure: The sub-6G and mmWave generation and their time-delay imposed} corresponds to spectrum \textcircled{4} and \textcircled{5} of Fig. \ref{Fig. A-RoF aided Multi-service Architecture}.}. Let us now consider a four-antenna-element system for both the sub-6GHz and mmWave transmission. The beating frequencies of the red boxes of Fig. \ref{Figure: The sub-6G and mmWave generation and their time-delay imposed} would be mapped to antenna elements 1-4 of both the mmWave antenna array and the sub-6GHz antenna array dispensing with the mmWave mixers and phase shifters, since they have been performed using the above optical link with the aid of WDM signal and CFBG.

Thus, a multi-service communication system transporting sub-6GHz and mmWave signals in the same direction is built for the two-hop HST systems. In the next section, we will detail the true-time delay principles and its mapping rule from the optical domain to the electronic domain as well as the multi-service generation.

\subsection{True-time Delay Principle}
\label{SubSection:True-time Delay Principle}
Instead of exploiting the conventional phase-shifting schemes for analogue beamforming \cite{5564376}, our proposed system invokes the true-time delay, which relies on the constant time-delay among the RF signal fed into the adjacent antennas elements to introduce the beam steering, which is capable of mitigating the wide-band beam-squinting problems \cite{cao2016advanced}. While undermining the codebook design of the phased-array systems \cite{7841766}, beam-squinting is also detrimental for the channel estimation and precoding, which further degrades the wireless communication transmission rate \cite{8882325}. 

In the following, with the aid of the mathematical derivation, we will prove how the imposed optical linear time-delay is used for introducing the constant time-delay among the RF signal fed into the adjacent antennas elements of each antenna array of Fig. \ref{Fig. A-RoF aided Multi-service Architecture}. In the upper line of the CU in Fig. \ref{Fig. A-RoF aided Multi-service Architecture}, where the spectrum \textcircled{1} is generated, the multi-service RF signals are used to directly modulate a laser diode and the input optical field of the MZM can be formulated as \cite{7055229}:
\begin{equation}
E_{in}(t)=\sqrt{P_{Laser}}e^{j\omega_{\lambda_1}t}[1+cos(\omega_{f_1}t)+cos(\omega_{f_2}t)+cos(\omega_{f_3}t)],
\label{Equation:Direct_modulation}
\end{equation}
where $P_{Laser}$ is the LD output power and $\omega_{f_1}$, $\omega_{f_2}$ and $\omega_{f_3}$ denote the optical carrier's angular frequency corresponding to $\lambda_1$, $\lambda_2$ and $\lambda_3$ of Fig. \ref{Fig. A-RoF aided Multi-service Architecture}. Then, the MZM output field \cite{7055229} is as follows:
\begin{equation*}
\begin{array}{lcl}
E_{MZM_{up}}(t)\\ = cos(\pm\frac{\pi}{4}+\frac{\pi V_{dr}cos(\omega_{\Delta f_{1}})}{2V_{\pi}})E_{in}(t)\\=\frac{\sqrt{P_{laser}}
[1+cos(\omega_{f_1}t)+cos(\omega_{f_2}t)+cos(\omega_{f_3}t)]e^{j\omega_{\lambda_1}t}}{\sqrt{2}}[J_{0}(\frac{\pi |V_{dr}|}{2V_{\pi}})\\+2\sum_{n=1}^{\infty}(-1)^nJ_{2n}(\frac{\pi V_{dr}}{2V_{\pi}})
cos(2n\omega_{\Delta f_{1}}t)\\\pm2\sum_{n=1}^{\infty}(-1)^nJ_{2n-1}(\frac{\pi V_{dr}}{2V_{\pi}})cos((2n-1)\omega_{\Delta f_{1}}t)]\\=\frac{\sqrt{P_{laser}}}{\sqrt{2}}[J_{0}(\frac{\pi |V_{dr}|}{2V_{\pi}})[e^{j\omega_{\lambda_1}t}+e^{j(\omega_{\lambda_1}+\omega_{f_1})t}/2\\+e^{j(\omega_{\lambda_1}t-\omega_{f_1}t)}/2+e^{j(\omega_{\lambda_1}+\omega_{f_2})t}/2+e^{j(\omega_{\lambda_1}t-\omega_{f_2}t)}/2\\+e^{j(\omega_{\lambda_1}+\omega_{f_3})t}/2+e^{j(\omega_{\lambda_1}t-\omega_{f_3}t)}/2]\\+2\sum_{n=1}^{\infty}(-1)^nJ_{2n}(\frac{\pi V_{dr}}{2V_{\pi}})\\\times[e^{j(\omega_{\lambda_1}+2n\omega_{\Delta f_{1}})t}/2+e^{j(\omega_{\lambda_1}-2n\omega_{\Delta f_{1}})t}/2\\+\frac{e^{j(\omega_{\lambda_1}+2n\omega_{\Delta f_{1}}+\omega_{f_1})t}/2+e^{j(\omega_{\lambda_1}-2n\omega_{\Delta f_{1}}-\omega_{f_1})t}/2}{2}\\+\frac{e^{j(\omega_{\lambda_1}+2n\omega_{\Delta f_{1}}-\omega_{f_1})t}/2+e^{j(\omega_{\lambda_1}-2n\omega_{\Delta f_{1}}+\omega_{f_1})t}/2}{2}\\+\frac{e^{j(\omega_{\lambda_1}+2n\omega_{\Delta f_{1}}+\omega_{f_2})t}/2+e^{j(\omega_{\lambda_1}-2n\omega_{\Delta f_{1}}-\omega_{f_2})t}/2}{2}\\+\frac{e^{j(\omega_{\lambda_1}+2n\omega_{\Delta f_{1}}-\omega_{f_2})t}/2+e^{j(\omega_{\lambda_1}-2n\omega_{\Delta f_{1}}+\omega_{f_2})t}/2}{2}\\+\frac{e^{j(\omega_{\lambda_1}+2n\omega_{\Delta f_{1}}+\omega_{f_3})t}/2+e^{j(\omega_{\lambda_1}-2n\omega_{\Delta f_{1}}-\omega_{f_3})t}/2}{2}\\+\frac{e^{j(\omega_{\lambda_1}+2n\omega_{\Delta f_{1}}-\omega_{f_3})t}/2+e^{j(\omega_{\lambda_1}-2n\omega_{\Delta f_{1}}+\omega_{f_3})t}/2}{2}]\\\pm2\sum_{n=1}^{\infty}(-1)^nJ_{2n-1}(\frac{\pi V_{dr}}{2V_{\pi}})\\\times[\frac{e^{j(\omega_{\lambda_1}+(2n-1)\omega_{\Delta f_{1}})t}+e^{j(\omega_{\lambda_1}-(2n-1)\omega_{\Delta f_{1}})t}}{2}\\+\frac{e^{j(\omega_{\lambda_1}+(2n-1)\omega_{\Delta f_{1}}+\omega_{f_1})t}/2+e^{j(\omega_{\lambda_1}-(2n-1)\omega_{\Delta f_{1}}-\omega_{f_1})t}/2}{2}\\+\frac{e^{j(\omega_{\lambda_1}+(2n-1)\omega_{\Delta f_{1}}-\omega_{f_1})t}/2+e^{j(\omega_{\lambda_1}-(2n-1)\omega_{\Delta f_{1}}+\omega_{f_1})t}/2}{2}\\+\frac{e^{j(\omega_{\lambda_1}+(2n-1)\omega_{\Delta f_{1}}+\omega_{f_2})t}/2+e^{j(\omega_{\lambda_1}-(2n-1)\omega_{\Delta f_{1}}-\omega_{f_2})t}/2}{2}\\+\frac{e^{j(\omega_{\lambda_1}+(2n-1)\omega_{\Delta f_{1}}-\omega_{f_2})t}/2+e^{j(\omega_{\lambda_1}-(2n-1)\omega_{\Delta f_{1}}+\omega_{f_2})t}/2}{2}\\+\frac{e^{j(\omega_{\lambda_1}+(2n-1)\omega_{\Delta f_{1}}+\omega_{f_3})t}/2+e^{j(\omega_{\lambda_1}-(2n-1)\omega_{\Delta f_{1}}-\omega_{f_3})t}/2}{2}\\+\frac{e^{j(\omega_{\lambda_1}+(2n-1)\omega_{\Delta f_{1}}-\omega_{f_3})t}/2+e^{j(\omega_{\lambda_1}-(2n-1)\omega_{\Delta f_{1}}+\omega_{f_3})t}/2}{2}]],
\label{Equation:MZM_multiple_signal_upper}
\end{array}
\end{equation*} 
where $\omega_{\Delta f_{1}}$, $V_{dr}$ and $V_{pi}$ are the angular frequency of $\Delta f_{1}$, the amplitude of the drive frequency of the MZM and its switching voltage. $J_{n}(\frac{\pi V_{dr}}{2V_{\pi}})$ is the Bessel function of the first kind and order n, which determines both the number and the amplitude of the side-bands. The spectrum of the MZM output is marked as \textcircled{1} of Fig. \ref{Fig. A-RoF aided Multi-service Architecture}, where the same RF signals subsuming $f_1$, $f_2$ and $f_3$ are carried by each wavelength, generating a WDM signal spaced with $\Delta f_1$. Similarly, the WDM signal at the output of the second MZM represented by spectrum \textcircled{2} in Fig. \ref{Fig. A-RoF aided Multi-service Architecture} can be expressed as follows, where the input optical field of the bottom MZM $E_{in2}(t)$ and an unmodulated WDM signal $E_{MZM_{bottom}}(t)$ is generated:
\begin{equation}
E_{in2}(t)=\sqrt{P_{Laser}}e^{j\omega_{\lambda_2}t},
\label{Equation:Direct_modulation2}
\end{equation}
\begin{equation}
\begin{array}{lcl}
E_{MZM_{bottom}}(t)\\ = cos(\pm\frac{\pi}{4}+\frac{\pi V_{dr}cos(\omega_{\Delta f_{2}})}{2V_{\pi}})E_{in2}(t)\\=  \frac{\sqrt{P_{laser}}
e^{j\omega_{\lambda_2}t}}{\sqrt{2}}[J_{0}(\frac{\pi V_{dr}}{2V_{\pi}})\\+2\sum_{n=1}^{\infty}(-1)^nJ_{2n}(\frac{\pi V_{dr}}{2V_{\pi}})
cos(2n\omega_{\Delta f_{2}}t)\\\pm2\sum_{n=1}^{\infty}(-1)^nJ_{2n-1}(\frac{\pi V_{dr}}{2V_{\pi}})cos((2n-1)\omega_{\Delta f_{2}}t)]\\=\frac{\sqrt{P_{laser}}}{\sqrt{2}}[J_{0}(\frac{\pi |V_{dr}|}{2V_{\pi}})(e^{j\omega_{\lambda_2}t})\\+2\sum_{n=1}^{\infty}(-1)^nJ_{2n}(\frac{\pi V_{dr}}{2V_{\pi}})\\\times[e^{j(\omega_{\lambda_2}+2n\omega_{\Delta f_{2}})t}/2+e^{j(\omega_{\lambda_2}-2n\omega_{\Delta f_{2}})t}/2]\\\pm2\sum_{n=1}^{\infty}(-1)^nJ_{2n-1}(\frac{\pi V_{dr}}{2V_{\pi}})\\\times[\frac{e^{j(\omega_{\lambda_2}+(2n-1)\omega_{\Delta f_{2}})t}+e^{j(\omega_{\lambda_2}-(2n-1)\omega_{\Delta f_{2}})t}}{2}]].
\end{array}
\end{equation}
\normalsize
Then, the two WDM signals are combined using the fiber coupler and transmitted through a single mode fiber. By assuming $\Delta f=\Delta f_1=\Delta f_2$ and $\Delta f=2(f_{02}-f_{01})$, the combined WDM signal has a wavelength spacing of $\frac{\Delta f}{2}.$ As shown in spectrum \textcircled{3} of Fig. \ref{Fig. A-RoF aided Multi-service Architecture}, the combined WDM signal would be filtered to two ports, with each being fed into a WDM Demux. Each output of the DWDM Demux would be fed into a photo detector to obtain either sub-6GHz signal or mmWave signal. Then, assuming the time delays imposed in the different frequencies are $t_n$ and $T_n$ as shown in Fig. \ref{Figure: The sub-6G and mmWave generation and their time-delay imposed} and by filtering the unwanted low frequency, we are capable of obtaining three frequencies in the mmWave spectrum as $S_{mmPD_1}$, $S_{mmPD_2}$ and $S_{mmPD_3}$ after the PD as shown in Fig. \ref{Fig. A-RoF aided Multi-service Architecture}, which are then input to a BPF and then EA\footnote{Here, we use three photo-detected signals as an example, which can be readily extended to any number of photo detectors.}. These mmWave signals can be represented as follows:

\tiny
\begin{equation}
\begin{array}{lcl}
S_{mmPD_1}=\frac{P_{laser}}{2}\\(cos((\omega_{\lambda_2}-\omega_{\lambda_1}+\omega_{{f_1}})(t+\frac{\omega_{\lambda_2}T_1-\omega_{{\Delta f}}T_1-\omega_{\lambda_1}t_3+\omega_{\Delta f}t_3+\omega_{f_1}t_3}{-(\omega_{\lambda_2}-\omega_{\lambda_1}+\omega_{{f_1}})}))+\\cos((\omega_{\lambda_2}-\omega_{\lambda_1}+\omega_{{f_2}})(t+\frac{\omega_{\lambda_2}T_1-\omega_{{\Delta f}}T_1-\omega_{\lambda_1}t_2+\omega_{\Delta f}t_2+\omega_{f_2}t_2}{-(\omega_{\lambda_2}-\omega_{\lambda_1}+\omega_{{f_2}})}))+\\cos((\omega_{\lambda_2}-\omega_{\lambda_1}+\omega_{{f_3}})(t+\frac{\omega_{\lambda_2}T_1-\omega_{{\Delta f}}T_1-\omega_{\lambda_1}t_1+\omega_{\Delta f}t_1+\omega_{f_3}t_1}{-(\omega_{\lambda_2}-\omega_{\lambda_1}+\omega_{{f_3}})})))
\end{array}
\label{Equation: mmPD_1}
\end{equation}

\begin{equation}
\begin{array}{lcl}
S_{mmPD_2}\\=\frac{P_{laser}}{2}(cos((\omega_{\lambda_2}-\omega_{\lambda_1}+\omega_{{f_1}})(t+\frac{\omega_{\lambda_2}T_2-\omega_{\lambda_1}t_3^{'}+\omega_{f_1}t_3^{'}}{-(\omega_{\lambda_2}-\omega_{\lambda_1}+\omega_{{f_1}})}))\\+cos((\omega_{\lambda_2}-\omega_{\lambda_1}+\omega_{{f_2}})(t+\frac{\omega_{\lambda_2}T_2-\omega_{\lambda_1}t_2^{'}+\omega_{f_2}t_2^{'}}{-(\omega_{\lambda_2}-\omega_{\lambda_1}+\omega_{{f_2}})}))\\+cos((\omega_{\lambda_2}-\omega_{\lambda_1}+\omega_{{f_3}})(t+\frac{\omega_{\lambda_2}T_2-\omega_{\lambda_1}t_1^{'}+\omega_{f_3}t_1^{'}}{-(\omega_{\lambda_2}-\omega_{\lambda_1}+\omega_{{f_3}})})))
\end{array}
\label{Equation: mmPD_2}
\end{equation}
\begin{table*}[h]
		\begin{center}
		\scalebox{1}{
		\begin{tabular}{l|c|c}
		\hline
		\hline Components & Proposed System & Conventional System \\
		\hline \multicolumn{3}{c}{Passive Components}\\
		 \hline Optical Interleaver & 1 & 0 \\
		 \hline CFBG & 1 & 0 \\
		    \hline DWDM Demux & 2 & 1 \\
		    \hline \multicolumn{3}{c}{Active Components}\\
		    \hline PD & 8 ($2N_A$) & 6 ($N_s$)\\	   
 \hline BPF & 8 ($2N_A$) & 24 ($N_sN_A$) \\
 \hline EA & 8 ($2N_A$)& 24 ($N_sN_A$)\\
 \hline Phase shifters & 0 & 12 mmWave PSs+12 Sub-6GHz PS ($N_sN_A$)\\
 \hline PSL & 0 & 6 ($N_s$)\\
		    \hline
		    \hline
		\end{tabular}
		}
	\end{center}
	\caption{RRH Complexity Comparison}
		\label{table:T1}
	\end{table*}
\begin{equation}
\begin{array}{lcl}
S_{mmPD_3}=\frac{P_{laser}}{2}\\(cos((\omega_{\lambda_2}-\omega_{\lambda_1}+\omega_{{f_1}})(t+\frac{\omega_{\lambda_2}T_3+\omega_{{\Delta f}}T_3-\omega_{\lambda_1}t_3^{''}-\omega_{\Delta f}t_3^{''}+\omega_{f_1}t_3^{''}}{-(\omega_{\lambda_2}-\omega_{\lambda_1}+\omega_{{f_1}})}))\\+cos((\omega_{\lambda_2}-\omega_{\lambda_1}+\omega_{{f_2}})(t+\frac{\omega_{\lambda_2}T_3+\omega_{{\Delta f}}T_3-\omega_{\lambda_1}t_2^{''}-\omega_{\Delta f}t_2^{''}+\omega_{f_2}t_2^{''}}{-(\omega_{\lambda_2}-\omega_{\lambda_1}+\omega_{{f_2}})}))\\+cos((\omega_{\lambda_2}-\omega_{\lambda_1}+\omega_{{f_3}})(t+\frac{\omega_{\lambda_2}T_3+\omega_{{\Delta f}}T_3-\omega_{\lambda_1}t_1^{''}-\omega_{\Delta f}t_1^{''}+\omega_{f_3}t_1^{''}}{-(\omega_{\lambda_2}-\omega_{\lambda_1}+\omega_{{f_3}})}))),
\end{array}
\label{Equation: mmPD_3}
\end{equation}
\normalsize

where $\omega_{mmWave_n}=\omega_{\lambda_2}-\omega_{\lambda_1}+\omega_{{f_n}}$ is the generated mmWave signal frequency. Then, if we take $\omega_{mmWave_1}$ as an example, according to Equations (\ref{Equation: mmPD_1}), (\ref{Equation: mmPD_2}) and (\ref{Equation: mmPD_3}), the time-delay difference between AE2 and AE1 can be referred to as $\Delta_{1}$, and between AE3 and AE2 as $\Delta_{2}$, which can be represented as:
\begin{equation}
\begin{array}{lcl}
\Delta_1=\frac{\omega_{\lambda_2}T_2-\omega_{\lambda_1}t_3^{'}+\omega_{f_1}t_3^{'}}{-(\omega_{\lambda_2}-\omega_{\lambda_1}+\omega_{{f_1}})}\\-\frac{\omega_{\lambda_2}T_1-\omega_{{\Delta f}}T_1-\omega_{\lambda_1}t_3+\omega_{\Delta f}t_3+\omega_{f_1}t_3}{-(\omega_{\lambda_2}-\omega_{\lambda_1}+\omega_{{f_1}})}\\=\frac{\omega_{mmWave_1}(T_2-T_1)+(n(T_2-T_1)/(2\pi))(2\pi\omega_{mmWave_1}/n)}{-\omega_{mmWave_1}}\\=-2(T_2-T_1),
\end{array}
\label{Equation A1A2td}
\end{equation}
and 
\begin{equation}
\begin{array}{lcl}
\Delta_2\\=\frac{\omega_{\lambda_2}(T_3-T_2)-\omega_{\lambda_1}(t_3^{''}-t_3^{'})+\omega_{f_1}(t_3{''}-t_3^{'})+\omega_{\Delta f}(T_3-t_3^{''})}{-(\omega_{\lambda_2}-\omega_{\lambda_1}+\omega_{{f_1}})}\\=-2(T_3-T_2).
\end{array}
\label{Equation A2A3td}
\end{equation}

As mentioned in Section \ref{Section:Introduction}, CFBG is capable of obtaining a linear relation between the time delay and the optical spectrum fed into it, which would be elaborated in Section \ref{Section:System Evaluation}. Then, we obtain $\Delta T=T_3-T_2=T_2-T_1$ as the time-delay difference of two frequencies with a spacing of $\Delta f$. Hence, we conclude that the optical time-delay of $\Delta T$ can result in a $-2\Delta T$ constant time-delay difference among the neighbouring AEs as verified by (\ref{Equation A1A2td}) and (\ref{Equation A2A3td}). This rule also applies to $\omega_{mmWave_1}$ and $\omega_{mmWave_2}$. Similarly, by repeating the process of the above derivation, we can have the time-delay difference relationship in terms of sub-6GHz as $\tau=-2\Delta T$.
Then, we attempt to solve the problem of the beam-squinting for the mmWave and the sub-6GHz signals, where we implement the CFBG3 of Fig. \ref{Fig. A-RoF aided Multi-service Architecture}, where the beam angle and the time delay relation is as follows:
\begin{equation}
\tau=\frac{dcos(\theta)}{c},
\end{equation}
where the $\theta$ is the beam angle to the perpendicular orientation of the antenna array. In order to obtain the same beam direction for both mmWave and sub-6GHz, the time delay difference of the sub-6GHz and the mmWave service must satisfy the following: $cos (\theta_1)=cos (\theta_2)=c\tau_1/d_1=c\tau_2/d_2$, where we have $\tau_2=\frac{d_2\tau_1}{d_1}$, with $d_1$ and $d_2$ being the inter-element distance of each antenna array. The CFBG3 of the RRH is capable of tuning the time delay of the mmWave service from $\tau_1$ to $\tau_2$. 

\begin{figure*}[h]
\centering
\includegraphics[width=.9\textwidth]{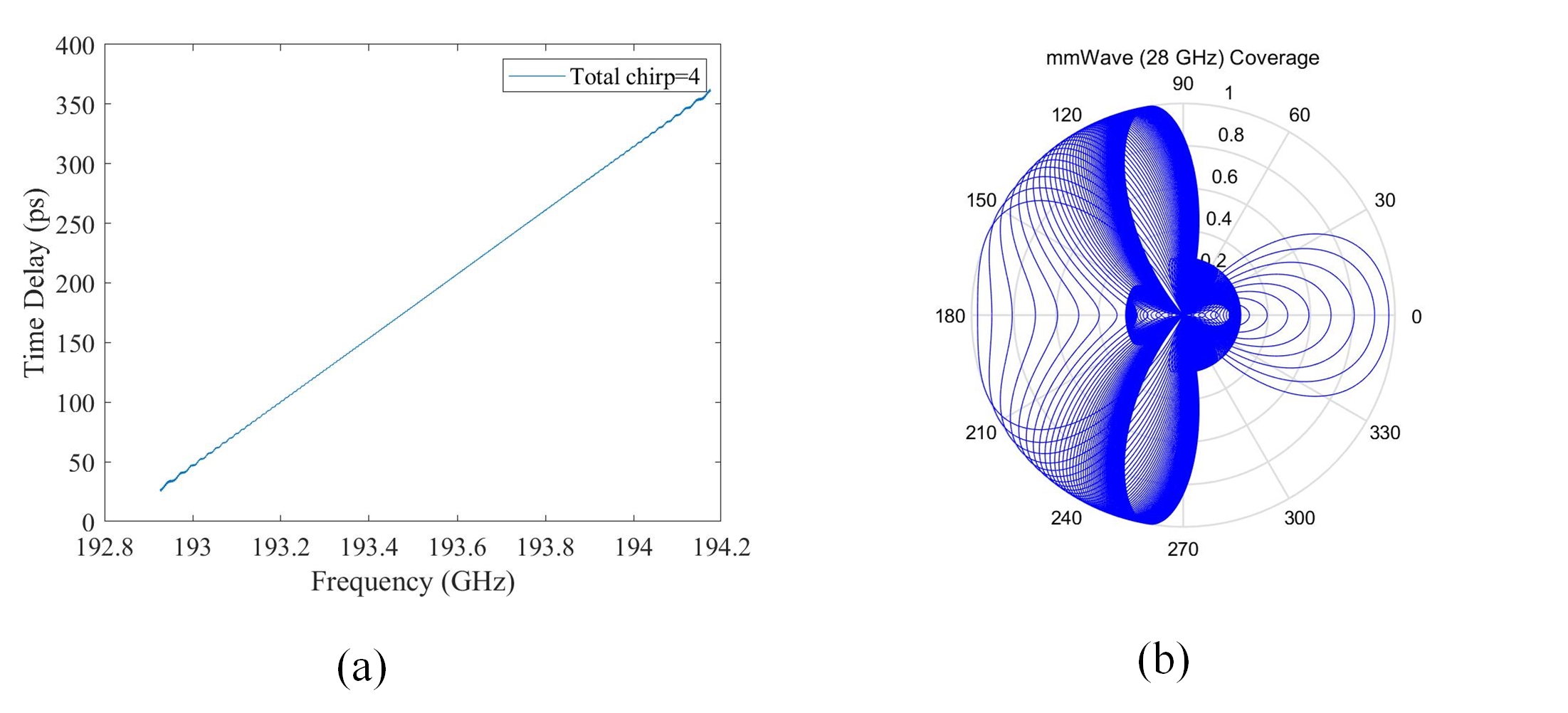} 
   \caption{The Time Delay and its Corresponding Beam coverage}
	\label{Figure:The Time Delay and its Corresponding Beam coverage}
\end{figure*}
\begin{figure*}[h]
\centering
\includegraphics[width=.9\textwidth]{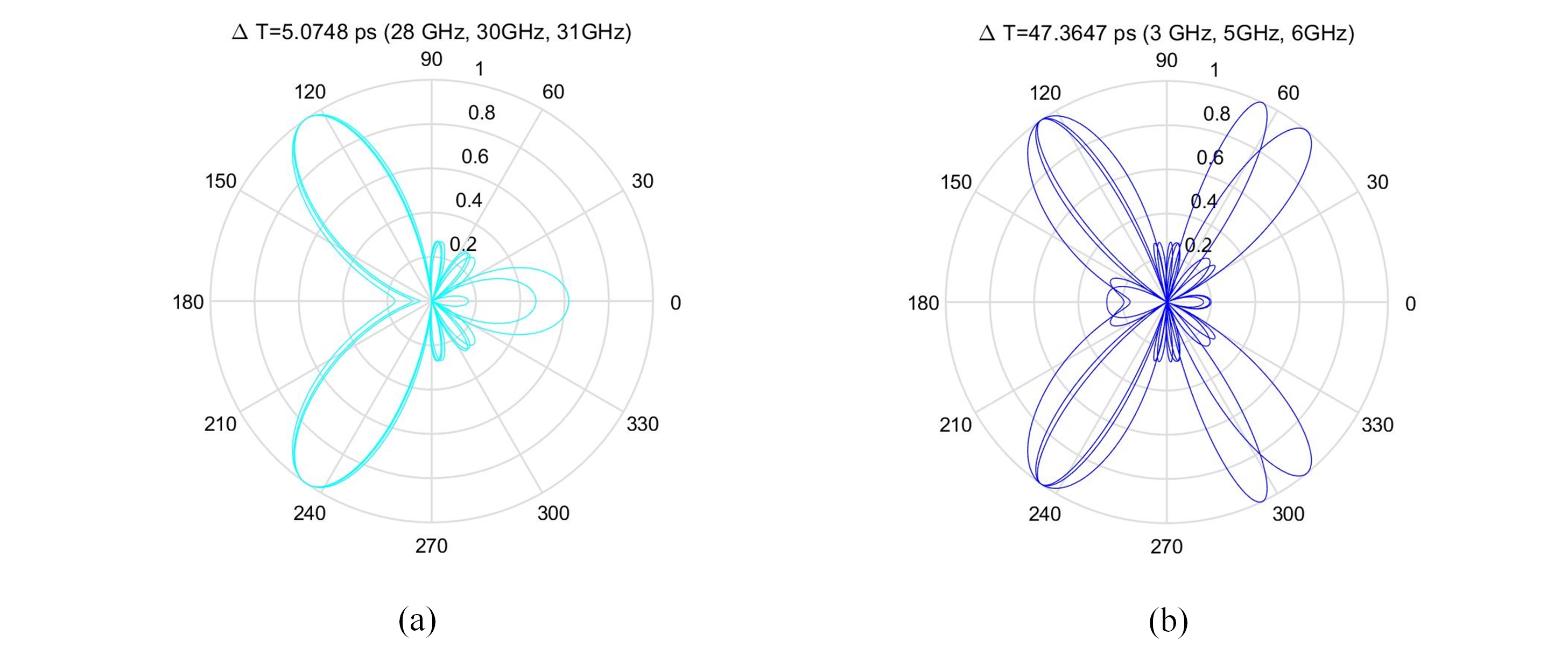}
   \caption{Multi-service Beam in the same direction. (a) The beam direction of mmWave signals at 28 GHz, 30 GHz and 31 GHz, when the true-time-delay $\Delta T$ of Section \ref{SubSection:True-time Delay Principle} is 5.0748. (b) The beam direction of sub-6GHz signals at 3 GHz, 5 GHz and 6 GHz, when the true-time-delay $\Delta T$ of Section \ref{SubSection:True-time Delay Principle} is 47.3647.}
	\label{Figure:Multi-service beam}
\end{figure*}
	\begin{table*}[h]
		\begin{center}
		\scalebox{1}{
		\begin{tabular}{cc}
		\hline
		\hline Simulation Parameters & Values \\
		 \hline Number of Antenna Element ($N_A$) & 4 \\
		 \hline Combined WDM Wavelength Spacing   & 25 GHz \\
		 \hline Length of the Chirped FBG & 40 mm\\
		    \hline  $f_1$, $f_2$ and $f_3$ & 3, 5, 6 GHz \\
		    \hline $f_{o1}$, $f_{o2}$ & 193.500, 193.525 GHz\\
		    \hline \multirow{2}{*}{WDM Central Frequencies} & 193.450, 193.475, 193.500, 193.525 GHz\\ & 193.550, 193.575, 193.600, 193.625 GHz\\
		    \hline RF signal Generation ($N_s$) & 3, 5, 6, 28, 30, 31 GHz\\
		    \hline Simulation Platform & Optisystem, OptiGrating \\	   
		    \hline
		    \hline
		\end{tabular}
		}
	\end{center}
	\caption{RRH Complexity Comparison}
		\label{table:T2}
	\end{table*}
\begin{figure*}[h]
\centering
\includegraphics[width=.9\textwidth]{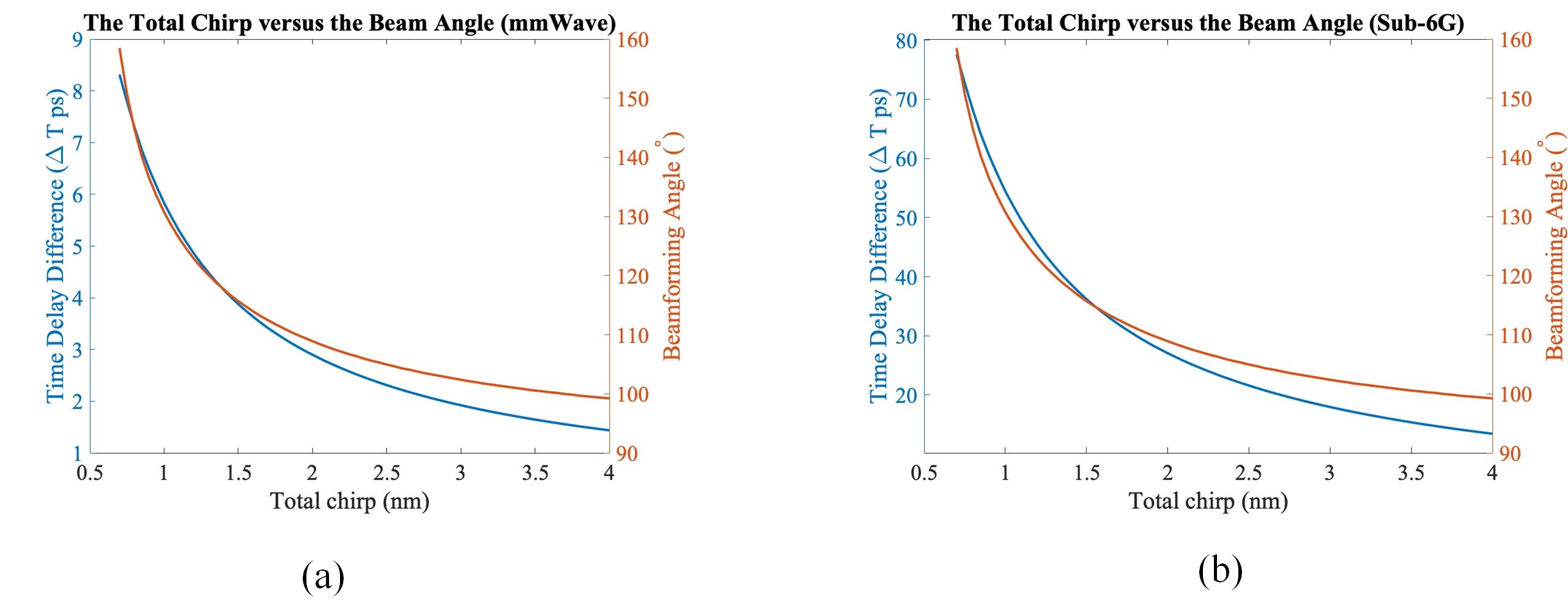}
   \caption{The total chirp versus the beam angle}
	\label{Figure:total chirp vs angle and the time-delay differenc}
\end{figure*}
Therefore, by assuming a linear relation of $t=nf+m$ between the optical frequency and the time delay imposed by the CFBG1 and CFBG2 of the CU of Fig \ref{Fig. A-RoF aided Multi-service Architecture}, where $n$ and $m$ are constants and the new linear relation imposed by the CFBG3 in the RRH is $t=fk+j$, we have a new relation of frequency and time-delay as $t=f(k+n)+m+j$ for the mmWave WDM signal. As mentioned above, we have the following relationship:
\begin{equation}
\tau_1=-2(T_2-T_1)=-2n\Delta f,
\end{equation}
\begin{equation}
\tau_2=-2(T_2^{new}-T_1^{new})=-2(n+k)\Delta f.
\end{equation}
Thus, $\frac{d_2\tau_1}{d_1}=-2(n+k)\Delta f$, and we obtain the relation $k=-\frac{d_2\tau_1}{2d_1\Delta f}-n=\frac{d_2-d_1}{d_1}n$. Then, by simply tuning the CFBG3 of the RRH according to the CFBG1 and CFBG2 of the CU of Fig. \ref{Fig. A-RoF aided Multi-service Architecture} to the new relation as $t=fk+j$, the multi-service communications including sub-6GHz and mmWave can be beamsteered to the same direction. Note that the linear relation can be tuned by changing the supported beam's deflection as demonstrated in \cite{1022008}, resulting in a tunable chirp rate of the linear CFBG, which is capable of obtaining a linear true-time delay of the WDM signals \cite{1022008}.

\textit{In this section, we verified mathematically that a double time-delay of the neighbouring wavelength of the WDM signal can be translated into the inter-element time-delay difference of the antenna array, justifying the feasibility of A-RoF aided beamforming control using optical true-time delays, while removing the electronic phase-shifters and avoiding the beam squints.} In the next section, we will evaluate our system in terms of  cost and  performance.

\section{System Evaluation}
\label{Section:System Evaluation} 
In this section, the system evaluation of our proposed system of Fig. \ref{Fig. A-RoF aided Multi-service Architecture} is presented. We will compare the cost and complexity of the proposed system with that of the conventional system of Fig. \ref{Figure: Conventional System}, following by a beamforming analysis of Fig. \ref{Fig. A-RoF aided Multi-service Architecture}.

\subsection{Cost and Complexity Analysis}
In this section, we analyse the cost benefits of our proposed design over the conventional system, where the multi-service bands are modulated using 6 LDs and beam-steered using 6 sets of phase-shifter array as shown in Fig. \ref{Figure: Conventional System}. As mentioned above, our design benefits from its ultralight RRH design, hence reducing the power-consumption and the OPEX. As shown in Table \ref{table:T1}, we list the number of components implemented in our proposed A-RoF system of Fig. \ref{Fig. A-RoF aided Multi-service Architecture} and the conventional system as shown in Fig. \ref{Figure: Conventional System}, where $N_S$ denotes the number of services while $N_A$ is the number of antenna elements of each antenna array. Assuming we transmit sub-6GHz spectrum of $f_1$, $f_2$ and $f_3$ and the mmWave frequencies of $f_1+\frac{\Delta f}{2}$, $f_2+\frac{\Delta f}{2}$ and $f_3+\frac{\Delta f}{2}$ using the four-antenna-element array, it is shown in Table \ref{table:T1} that in the proposed design of Fig \ref{Fig. A-RoF aided Multi-service Architecture}, in contrast to the conventional design of Fig. \ref{Figure: Conventional System}, the bulky phase-shifters and the power splitter are totally removed, while the number of BPF, EA is substantially reduced. Instead, some passive components such as optical interleaver, CFBG and an extra DWDM Demux are introduced, but the number of which is not dependent on either the services or the antenna elements. Explicitly, as seen in Table \ref{table:T1}, regarding the active components, the power-thirsty phase-shifter and power splitters are reduced from 24 and 6 to 0 and 0, respectively, while the number of BPFs and EAs are decreased from 24 and 24 to 8 and 8, respectively, resulting in a reduced power-comsumption in the RRH of Fig. \ref{Fig. A-RoF aided Multi-service Architecture}. Therefore, we can conclude that the power-consumption can be substantially reduced with the removal of the phase-shifters and the introduction of the passive optical components, while the total hardware cost would be reduced due to the optical mmWave generation without using the mmWave phase-shifters, simplifying the BPF and the EA. 

\subsection{Beamforming Performance Analysis}
\label{subsection: Beamforming Performance Analysis}
Let us now analyse the system performance of our proposed system. In this section, we use the uniform linear antenna array having four elements\footnote{Note that any number can be used and the four-element system is just an example}. The simulation parameters are listed in Table \ref{table:T2}. We directly modulate the laser diode using different frequencies of $f_1=3$, $f_2=5$ and $f_3=6$ GHz. Then, after the true-time-delay process and the optical interleaver, a photonic RF generation of sub-6GHz at  $f_1=3$, $f_2=5$ and $f_3=6$ GHz and mmWave at $f_1+\frac{\Delta f}{2}=28$, $f_2+\frac{\Delta f}{2}=30$ and $f_3+\frac{\Delta f}{2}=31$ GHz can be obtained without the bulky and RF mixers, where $\Delta f=\Delta f_1=\Delta f_2$. 

In this section, we present our novel beamforming design, where we obtain a relation between the optical signal after the CFBG1 and CFBG2 in the CU of Fig. \ref{Figure: The sub-6G and mmWave generation and their time-delay imposed} and its time delay imposed in Fig. \ref{Figure:The Time Delay and its Corresponding Beam coverage}(a). Explicitly, as portrayed in Fig. \ref{Figure:The Time Delay and its Corresponding Beam coverage} by tuning a supported beam's deflection as detailed in \cite{1022008}, we are capable of changing the total chirp, resulting in a linear relation between time-delay and the optical frequency as depicted in Fig. \ref{Figure:The Time Delay and its Corresponding Beam coverage}(a), hence resulting in a multi-service beam coverage around 180$^\circ$ as shown in Fig. \ref{Figure:The Time Delay and its Corresponding Beam coverage}(b). 
	
Moreover, to prove that our design can provide the same directive beam for both sub-6GHz and mmWave spectrum, as analysed in Section \ref{Section:A-RoF aided Beamforming for High-speed Train Multi-service Communications}, the CFBG1 and CFBG2 of Fig. \ref{Fig. A-RoF aided Multi-service Architecture} are required to impose the same degree of time-delay difference, while CFBG 3 is tuned to a time-delay related to the CFBG 1 and CFBG 2. As shown in Fig \ref{Figure:Multi-service beam}, the sub-6GHz beams are directed to the same direction at $\Delta T_1=47.3647$ to the mmWave beam at $\Delta T_2=5.0748$, where the relation happens to be  $\Delta T_1=d_2*\Delta T_2/d_1$ \footnote{In our example, $d_1=\lambda_{3GHz}/2$ and $d_2=\lambda_{28GHz}/2$ are the inter-element distances of the sub-6GHz antenna array and mmWave antenna array, respectively, where we assume the center frequencies are 3 GHz and 28 GHz. Then we have $d_2/d_1=0.107$.}. 

Explicitly, it is shown in Fig. \ref{Figure:total chirp vs angle and the time-delay differenc}(a) and \ref{Figure:total chirp vs angle and the time-delay differenc}(b) that the sub-6GHz and mmWave band are capable of being mapped to the same beamforming angle, thanks to the introduction of CFBG 3 in the RRH of Fig. \ref{Fig. A-RoF aided Multi-service Architecture}. When the total chirp of both the CFBG 1 and CFBG 2 ranges from 0.7 to 4 nm with a step-size of 0.05, the time delay difference of sub-6GHz and mmWave spans from 77.6 ps to 13.4 ps and from 8.3 to 1.4 ps, respectively, which results in a beam angle from 158.48 to 99.23 $^\circ$ as shown in Fig. \ref{Figure:total chirp vs angle and the time-delay differenc}(a) and \ref{Figure:total chirp vs angle and the time-delay differenc}(b). Thus, by controlling the total chirp of the CFBG 3 in the RRH according to the total chirp in the CU, we can direct the multi-service to the same direction as shown in Fig. \ref{Figure:Multi-service beam}.

Then, as shown in Fig. \ref{Figure:Multi-service beam}(b) and Fig. \ref{Figure:The Time Delay and its Corresponding Beam coverage}, the corresponding multi-service beam coverage, which is around 180$^\circ$ with a high precision, verifies that our photonic beamforming system can radiate the wideband beam flexibly and widely, while substantially simplifying the RRH designs as analysed in Section \ref{Section:A-RoF aided Beamforming for High-speed Train Multi-service Communications}.

\section{Conclusion}
\label{subsection: Conclusion}
In this article, we have proposed a low-cost A-RoF aided multi-service communications in the two-hop relay train system, where we implemented the photonic beamformer using the CFBGs, enabling a centralised and low-cost RAN design. In this paper, we designed a double time-delay mapping rule of the optical signals and the corresponding RF signals and verified that the proposed system was capable of reducing the total cost of the RAN by simplifying the RRHs. Finally, a single beam transmitting the multi-service signals with a 180$^\circ$ beamfroming range was presented.
\bibliographystyle{ieeetr}
\bibliography{ECS}
\begin{IEEEbiography}[{\includegraphics[width=1in,height=1.7in,clip,keepaspectratio]{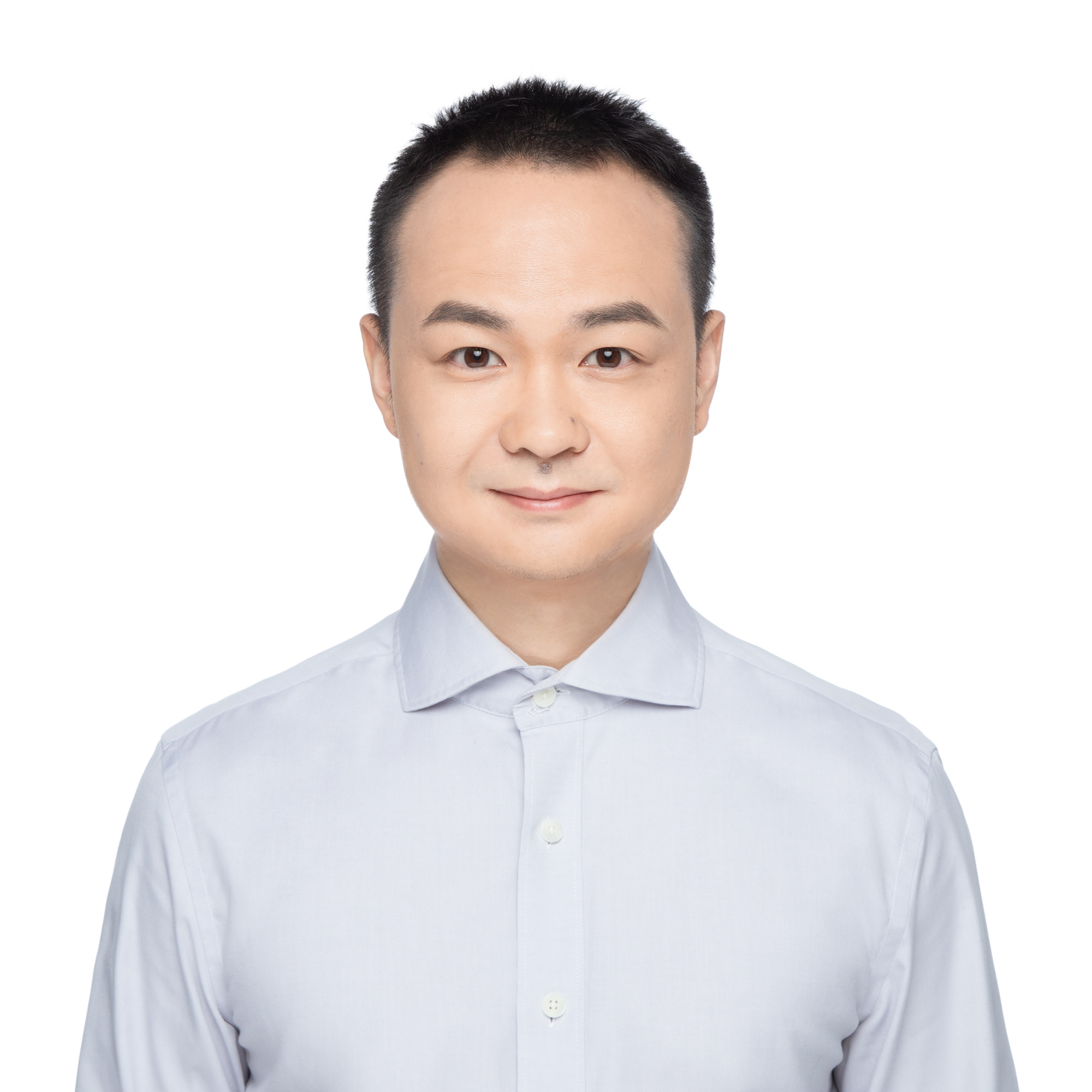}}]{Yichuan Li}
is an Assistant Professor with Harbin Institute of Technology (Shenzhen), Shenzhen, China.
He received B.Sc. degree in Optics Information
Science and Technology from China University
of Petroleum (East China), Qingdao, China, in
2012, and M.Sc. degree in wireless communications
from the University of Southampton, Southampton,
UK., in 2014. He was a research assistant in the Lightwave Communication Lab of the Chinese University of Hong Kong (CUHK) from July to October in 2017. He received his Ph.D. degree in wireless communications
from the University of Southampton, Southampton,
UK. in 2019.
His research is focused on the the radio over fiber
for backhaul, fronthaul and indoor communication
network. His research interests are millimeter wave
over fiber, optical fiber aided analogue beamforming
techniques, Multi-functional MIMO, mode division multiplexing in multimode
fiber and fiber-based C-RAN system.
\end{IEEEbiography}

\begin{IEEEbiography}[{\includegraphics[width=1in,height=1.25in,clip,keepaspectratio]{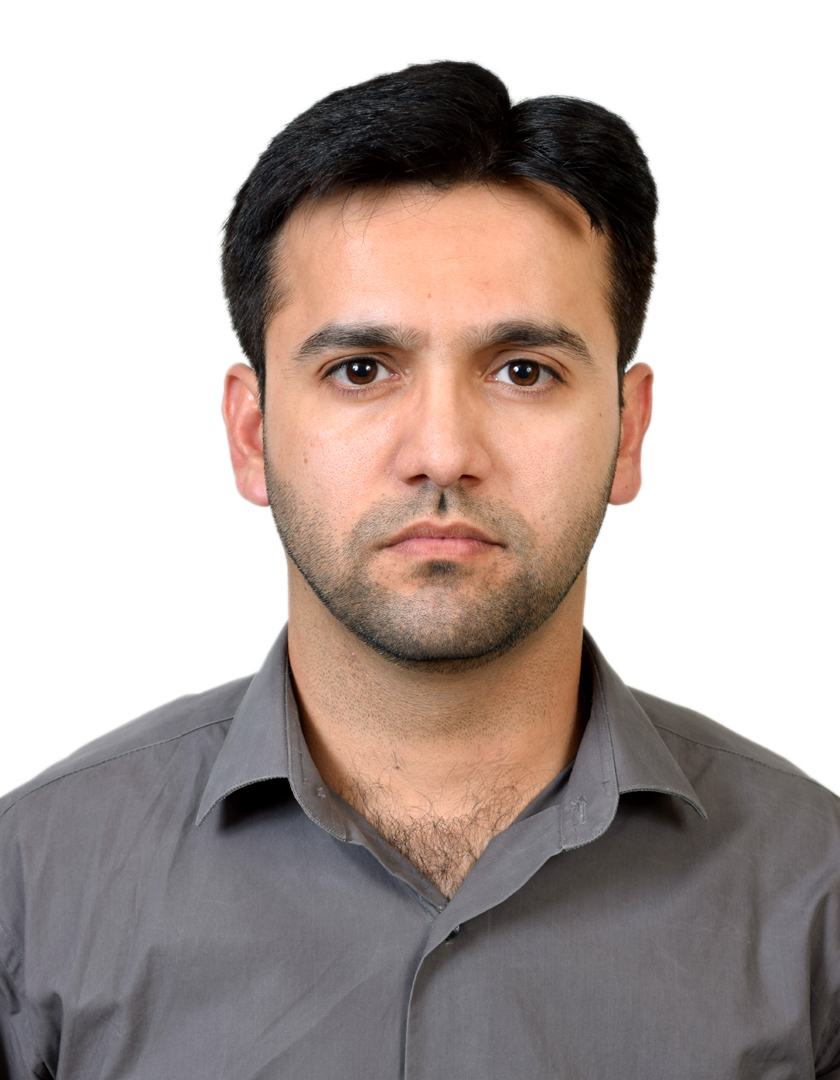}}]{Salman Ghafoor}
received the B.Sc. Electrical Engineering degree from UET Peshawar, Peshawar, Pakistan, in 2006, the M.Sc. degree in electronic communications and computer engineering from the University of Nottingham, Nottingham, U.K, and the Ph.D. degree from the University of Southampton, Southampton, U.K. He was a Research Student at the Optoelectronics Research Centre (ORC), University of Southampton, U.K for two years. In 2010, he joined the School of Electronics and Computer Science (ECS), University of Southampton where he completed his Ph.D in 2012. He is currently an Associate Professor with National University of Sciences and Technology (NUST), Islamabad, Pakistan. His areas of research include Free space optical communications, all-optical signal processing, ultra-wideband over fibre, and radio over fibre systems.
\end{IEEEbiography}

\begin{IEEEbiography}[{\includegraphics[width=1in,height=1.25in,clip,keepaspectratio]{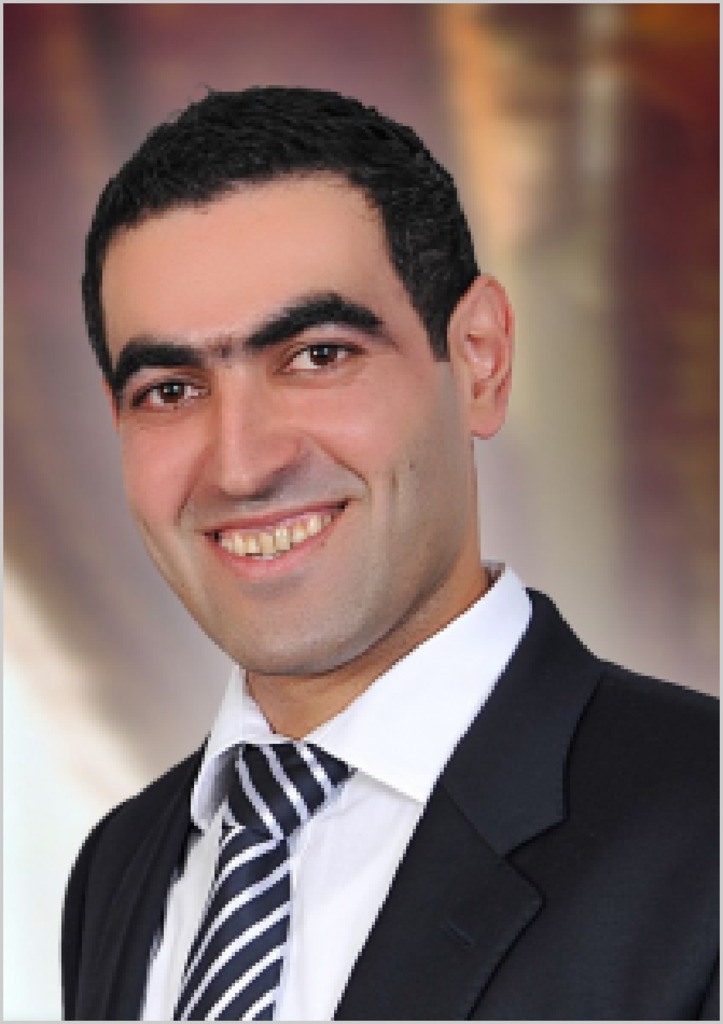}}]{Mohammed El-Hajjar} is an Associate Professor in the School of Electronics and Computer Science in the University of Southampton. He received his PhD in Wireless Communications from the University of Southampton, UK in 2008. Following the PhD, he joined Imagination Technologies as a design engineer, where he worked on designing and developing Imagination's multi-standard communications platform, which resulted in three patents. He is the recipient of several academic awards and has published a Wiley-IEEE book and in excess of 80 journal and conference papers. Mohammed's research interests include the development of intelligent communications systems, energy-efficient transceiver design, MIMO, millimeter wave communications and Radio over fiber network design.
 \end{IEEEbiography}

\end{document}